\documentclass[12pt]{article}\usepackage[hyperfootnotes=false]{hyperref}
\usepackage{epsfig}
\usepackage{float}

\usepackage{caption}
\usepackage{soul}
\usepackage{physics}
\usepackage{amsmath}
\usepackage{amssymb}
\usepackage{graphicx}
\newlength{\abstractwidth}
\renewcommand{\thefootnote}{\fnsymbol{footnote}}
\renewcommand{\thanks}[1]{\footnote{#1}} 
\newcommand{\starttext}{
\setcounter{footnote}{0}
\renewcommand{\thefootnote}{\arabic{footnote}}}

\newcommand{\be}{\begin{equation}}
\newcommand{\bea}{\begin{eqnarray}}
\newcommand{\eea}{\end{eqnarray}}
\newcommand{\beq}{\begin{equation}}
\newcommand{\ee}{\end{equation}}

\def\simleq{\; \raise0.3ex\hbox{$<$\kern-0.75em
\raise-1.1ex\hbox{$\sim$}}\; }
\def\simgeq{\; \raise0.3ex\hbox{$>$\kern-0.75em
\raise-1.1ex\hbox{$\sim$}}\; }

\def\bi{\begin{itemize}}
\def\ei{\end{itemize}}

\def\bn{\bigskip \noindent}

\makeatletter
\g@addto@macro\normalsize{%
  \setlength\abovedisplayskip{10pt}
  \setlength\belowdisplayskip{20pt}
  \setlength\abovedisplayshortskip{10pt}
  \setlength\belowdisplayshortskip{20pt}
}
\makeatother

\usepackage{color}

\begin{document}
  
\begin{titlepage}

\rightline{}
\bigskip
\bigskip\bigskip\bigskip\bigskip
\bigskip

\centerline{\Large \bf {The Case of the Missing Gates:}  } \vspace{3mm} 
\centerline{\Large \bf { Complexity of Jackiw-Teitelboim Gravity}}
\bn

\bigskip
\begin{center}
\bf   Adam R.~Brown$^{1,2}$, Hrant Gharibyan$^1$, Henry W. Lin$^3$,   \\ 
Leonard Susskind$^1$, L\'arus Thorlacius$^{1,4}$, Ying Zhao$^{1,5}$ \rm

\bigskip
$^1$ Stanford Institute for Theoretical Physics and Department of Physics, \\
Stanford University,
Stanford, CA 94305, USA \\

\bigskip
$^2$ Google, Mountain View, CA 94043, USA 

\bigskip
$^3$ Department of Physics, Princeton University, Princeton, NJ 08540, USA


\bigskip
$^4$ University of Iceland, Science Institute, Dunhaga 3, 107 Reykjavik, Iceland

and

The Oskar Klein Centre for Cosmoparticle Physics, Department of Physics, 
Stockholm University, AlbaNova, 106 91 Stockholm, Sweden

\bigskip
$^5$ Institute for Advanced Study, Princeton, NJ 08540, USA

\end{center}

\bn

\begin{abstract}

The Jackiw-Teitelboim (JT) model arises from the dimensional reduction of charged 
black holes.  Motivated by the holographic complexity conjecture, we calculate the 
late-time rate of change of action of a Wheeler-DeWitt patch in the JT theory. Surprisingly, 
the rate vanishes. This is puzzling because it contradicts both holographic expectations 
for the rate of complexification and also action calculations for charged black 
holes.  We trace the discrepancy to an improper treatment of boundary terms when 
naively doing the dimensional reduction. Once the boundary term is corrected, we find 
exact agreement with expectations. We comment on the general lessons that this might 
hold for holographic complexity and beyond. 

\bn
 
\end{abstract}

\end{titlepage}

\starttext \baselineskip=17.63pt \setcounter{footnote}{0}
\tableofcontents

\section{Introduction} \label{sec: intro}

The Jackiw-Teitelboim (JT) model of 1+1-dimensional 
dilaton gravity \cite{Jackiw:1984je,Teitelboim:1983ux}
is useful for studying conjectures relating the geometry of black holes and 
scrambling in dual quantum systems. This simple model holographically reproduces the 
nearly conformal dynamics \cite{nads2} of the Sachdev-Ye-Kitaev (SYK) 
model \cite{Sachdev:1992fk,Kitaev} at low energies. In the present paper, we will focus on a 
particular aspect of the duality---the conjectured relationship between quantum complexity and emergent spacetime. 

The SYK model is a good place to test the conjectured relation between complexity and the 
size of wormholes \cite{Susskind:2014rva,Brown:2015bva,Brown:2015lvg}. For one thing the SYK model is, from the beginning, 
a theory of (fermionic) qubits, which means that in principle the definition of complexity could be fairly standard. 
On the other hand, the gravitational dual at sufficiently low energies is fairly well understood.

In particular, the low-energy dynamics of the SYK model is characterized by a spontaneously and explicitly 
broken reparameterization symmetry, with the effective action of the associated Nambu-Goldstone modes 
given by the Schwarzian \cite{Almheiri:2014cka,Engelsoy:2016xyb,Maldacena:2016hyu}.
The resulting universal dynamics exhibits a number of distinctive features, including out-of-time-order
four point functions that saturate the chaos bound \cite{Maldacena:2015waa}. The boundary 
dynamics of 1+1-dimensional JT dilaton gravity is governed by the same broken time-reparameterization
symmetry and Schwarzian action, motivating its identification as a bulk dual for low-energy
SYK dynamics. At the same time, JT theory is simple enough to allow explicit computations 
of both the volume of a maximal surface as well as the action of a Wheeler-DeWitt (WDW) patch. 
This gives us the opportunity to test both the complexity-volume (CV) duality and
complexity-action (CA) duality against expectations from the quantum-chaotic nature of the 
SYK model.

For most purposes, JT theory is the dimensional reduction of the near-horizon dynamics of a
near-extremal Reissner-Nordstr\"om (RN) black hole of Einstein-Maxwell theory, and one might 
think they give identical results for holographic complexity. 
Indeed for CV duality JT and RN agree with one another, and with the expected 
behavior of complexity for SYK quantum mechanics. 
Surprisingly, when we compute the action in WDW patches we do not find agreement. 
The late-time rate of growth of action for the JT theory vanishes, in disagreement with our 
expectations for the growth of complexity for chaotic systems like SYK.\footnote{Throughout
this paper, ``late-time'' refers to much longer than the thermal time scale but less 
than the exponential time scale on which complexity is expected to saturate and on which 
classical geometry no longer provides a reliable description~\cite{Susskind:2014rva}.}

At first sight, this seems to be a serious counterexample to the CA conjecture.
On the other hand when the Einstein-Maxwell action is computed for the near-extremal 
RN black hole, we find perfect agreement with the expectations for complexity growth.
By carefully constructing the dimensional reduction of RN we have been able to trace the origin of the 
discrepancy. In eliminating the gauge field from RN, the dimensional reduction inadvertently introduces 
an inappropriate boundary term at the edge of the WDW patch. There does not seem to be a consistent 
way to remove this term without bringing back the gauge field as a degree of freedom. 

We emphasize that for the purposes for which it has been used up till now
the JT model correctly reflects much of the physics of the SYK model, 
including the low-energy dynamics governed by broken reparametrization 
symmetry at low temperature. We find, however, that for more subtle quantities, 
such as holographic complexity, the standard formulation of the JT model as a 
theory of two-dimensional dilaton gravity
fails to give sensible results, whereas
the original higher-dimensional Einstein-Maxwell theory succeeds.

The paper is organized as follows. In Section~\ref{sec:RN_C=A} we set the stage for the rest of 
the discussion by reviewing salient features of the near-extremal limit of electrically charged
RN black holes in 3+1-dimensional Einstein-Maxwell theory. In Section~\ref{JTmodel} we
write down the action of the JT model and describe its black hole solutions. 
This serves to fix notation and set up a simple 1+1-dimensional framework where explicit
calculations can be carried out to test proposals for holographic complexity. 
CV duality for JT black holes is considered in Section~\ref{sec:C=V} and found to 
reproduce predictions for higher-dimensional near-extremal RN black holes. In Section~\ref{sec:CvsA} we turn our attention to CA duality and
find that it apparently runs into a problem in the JT theory. In Section~\ref{sec:JTmodel} we
take steps to resolve the problem by re-examining how JT gravity arises as an effective theory 
for low-energy radial modes in the near-horizon region of a near-extremal RN black hole.
The eventual resolution involves a careful treatment of boundary terms in the action and is 
presented in Section~\ref{sec:theresolution}. Further examples of systems where boundary 
terms play similar roles are presented in Appendices. \\

Note added: the conclusions reached in this paper have also been independently reached by Goto, Marrochio, Myers, Queimada \& Yoshida, in a paper that will appear soon \cite{HugoEtAl}.

\section{Complexity = Action for RN Black Holes}  \label{sec:RN_C=A}

We begin our discussion by recalling some facts about charged black holes
in 3+1-dimensional Einstein-Maxwell theory with action
\begin{equation}
\mathcal{S}=\frac{1}{16\pi}\int_\mathcal{M} d^4x\,\sqrt{-g}\,\left(\frac{1}{\ell^2}R- \,
F_{\mu\nu}F^{\mu\nu}\right)+
\frac{1}{8\pi \ell^2}\int_\mathcal{\partial M} d^3y\,\sqrt{-h}\,(K-K_0)\,,
\label{4d_action}
\end{equation}
where $\ell \equiv \sqrt{G_N}$ is the 3+1-dimensional Planck length. 
This action describes an electromagnetic 
field gravitationally back-reacting on curved spacetime. 
It includes the usual Gibbons-Hawking-York 
boundary term \cite{Gibbons:1976ue,York:1972sj} involving the trace of the extrinsic curvature 
$K$ at an asymptotic spacetime boundary with induced metric $h_{ij}$. It also includes a regulator 
term that subtracts $K_0$, the trace of the extrinsic curvature of the same boundary surface 
when embedded in a flat spacetime, in order to obtain a finite free energy from the 
corresponding on-shell Euclidean action.\footnote{Alternatively, one can introduce a small
negative cosmological constant into the 3+1-dimensional theory and include the standard 
boundary counterterms that render the free energy finite.} 

The boundary conditions obeyed by the electromagnetic
field at $\partial\mathcal{M}$ will play an important role in our discussion. As it stands, the
action \eqref{4d_action} does not include any boundary term involving the Maxwell field and 
$A_\mu$ is kept fixed at the boundary. In the Euclidean formalism this corresponds to a thermal 
ensemble where the chemical potential is held fixed but the total electric charge of the system is 
allowed to fluctuate. If, on the other hand, the following boundary term is added to the action, 
\begin{equation}
\mathcal{S}_b^\textrm{em} 
= \frac{1}{4\pi}\int_\mathcal{\partial M} d^3y\,\sqrt{-h}\,\hat{n}_\mu\, F^{\mu\nu}A_\nu  \,,
\label{maxwellboundaryterm}
\end{equation}
then free variations of $A_\mu$ are allowed at the boundary and the corresponding 
thermal ensemble is that of fixed charge but varying chemical potential. 
We will revisit 3+1-dimensional Maxwell boundary terms in Appendix~\ref{sec:Maxwellboundaryterm}. 

A notable solution for the action \eqref{4d_action} is the spherically symmetric
Reissner-Nordstr\"om black hole with electric charge $Q>0$ and mass
$M\geq Q/\ell$, described by
\begin{eqnarray}
ds^2 &=& -f(r) dt^2+–\frac{dr^2}{f(r)}+r^2 d\Omega^2 \,, \nonumber\\
f(r)&=&\left(1-\frac{r_+}{r}\right)\left(1-\frac{r_-}{r}\right), \label{RNsolution} \\
F_{rt}&=&\frac{Q}{r^2}\,, \nonumber
\end{eqnarray}
where $r_\pm = \ell^2 M\pm\sqrt{\ell^4 M^2-\ell^2 Q^2}$ are the locations of the 
outer and inner horizon.  

We will be interested in these black holes when they are very near to 
extremality,  $M\rightarrow M_0=Q/\ell$. In the extremal limit, the  horizons are 
degenerate $r_+=r_-=\ell Q$ and the Hawking temperature,
\begin{equation}
T = \frac{(r_+-r_-)}{4\pi r_+^2} \,,
\label{Htemperature}
\end{equation}
goes to zero. Thus for near-extremal black holes $r_+-r_-\ll r_+$, or equivalently $\beta \gg r_+$.  
Following our recent work \cite{Brown:2018kvn}, we find it useful
to divide the spacetime geometry outside a near-extremal black hole into 
the three regions shown in Figure~\ref{fig:RNthroat}:

\begin{figure}[htbp] 
   \centering
   \includegraphics[width=4in]{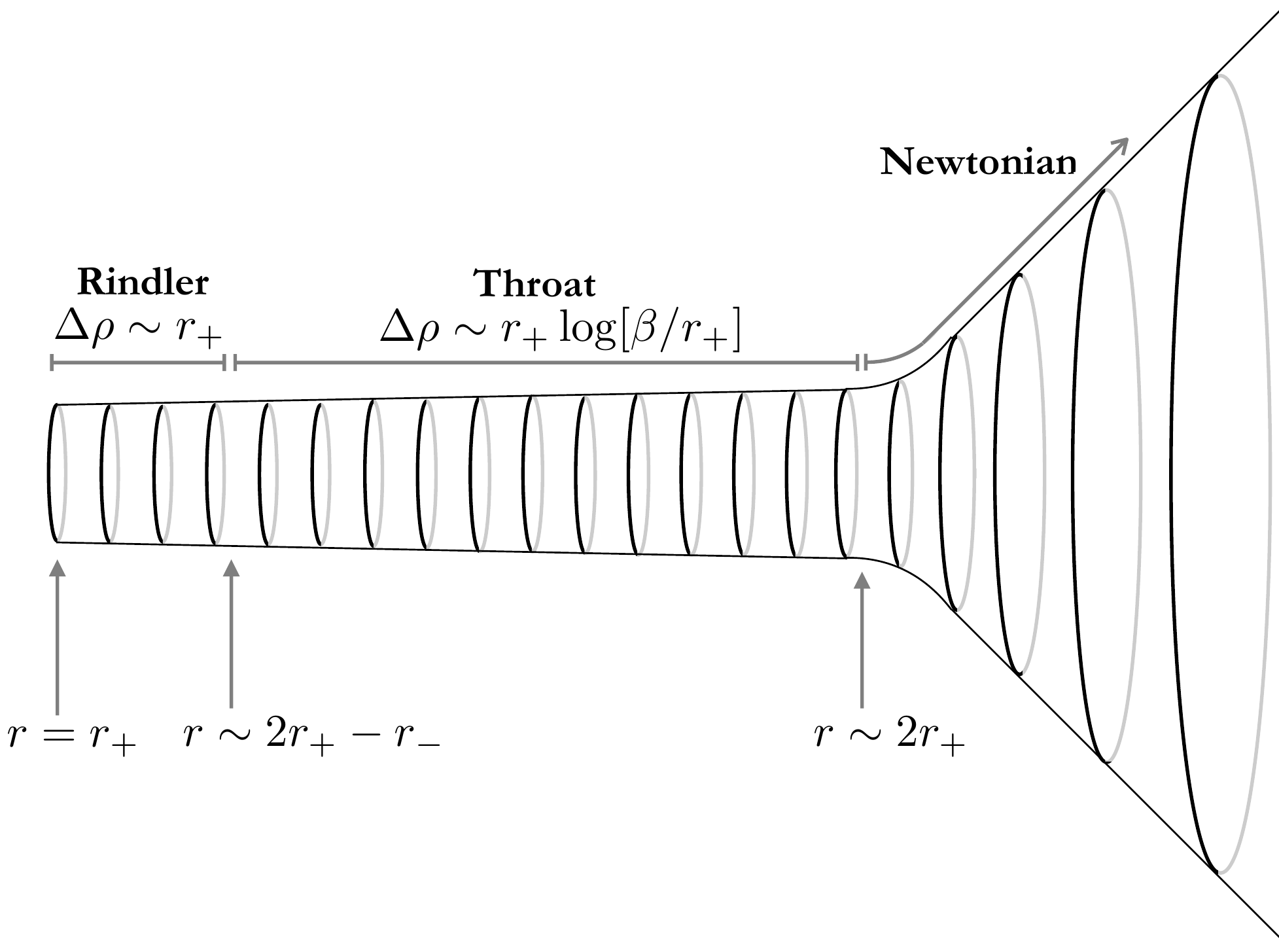} 
 \caption{The three regions outside the horizon of a near-extremal RN black hole. 
 The transition from throat to Newtonian region occurs at $r \sim 2r_+$, which is also 
 the approximate location of the top of the potential barrier \cite{Brown:2018kvn}, 
 and also the approximate location of the curved JT boundary in Fig.~\ref{fig:WDW_patch}.}
   \label{fig:RNthroat}
\end{figure}
\begin{itemize}
\item Closest to the outer horizon of the black hole is the Rindler region, $r_+\> < \> r \> \lesssim \> 2r_+-r_-$,
where the proper distance from the horizon, $\Delta\rho= \int_{r_+}^r dr'/\sqrt{f(r')}$, is in the range
\begin{equation}
0\> < \> \Delta\rho \> \lesssim \> r_+ \,.
\end{equation} 
\item
The ``throat'' region, $2r_+-r_- \> \lesssim \> r \> \lesssim \> 2r_+$, 
becomes long in the low-temperature limit, 
\begin{equation}
r_+\> \lesssim \> \Delta\rho \> \lesssim \> r_+ \log\left(\frac{\beta}{r_+}\right) \,,
\end{equation} 
and in this case the spacetime geometry is well approximated locally by AdS$_2\times S^2$.
The long throat supports a low-energy sector of long-wavelength 
radial excitations that are described by an effective two-dimensional theory, 
the Jackiw-Teitelboim model, discussed in Section~\ref{JTmodel} below.
\item
Finally, there is the Newtonian region, $2r_+ \> \lesssim \> r $, where the metric 
approaches that of flat spacetime $f(r)\approx 1$. The JT boundary from Fig.~\ref{fig:WDW_patch} 
may be identified with where the AdS$_2$ throat meets the Newtonian region. 
\end{itemize}

In \cite{Brown:2015bva,Brown:2015lvg}, the late-time rate of change of the action of a WDW patch for RN black 
holes was calculated to be 
\begin{equation}
\frac{dS}{d (t_L + r_R) }   = \frac{Q^2 }{r_-} - \frac{Q^2 }{r_+} =4\, S\, T \,, 
\label{eq:chargedincrement}
 \end{equation}
where $T$ is the Hawking temperature \eqref{Htemperature} and $S=\pi r_+^2/\ell^2$ is the 
Bekenstein-Hawking entropy. 
For our purposes, the essential aspect of this result is that the action advances linearly at late times. 
This is consistent with expectations for complexity---at pre-exponential times, a generic quantum 
system will complexify with constant rate. However, it is superficially inconsistent 
with the vanishing rate of action advance that we obtain for the JT theory in Sec.~\ref{sec:CvsA}. 
This inconsistency will be resolved in Sec.~\ref{sec:theresolution}, in favor of Eq.~\ref{eq:chargedincrement}.

\section{Jackiw-Teitelboim model}
\label{JTmodel}

We now turn to our main object of study, the JT model of 1+1-dimensional 
dilaton gravity \cite{Jackiw:1984je,Teitelboim:1983ux}.
We start by introducing the basic equations of the model and write down a family of static solutions. 
This serves to establish notation\footnote{For the most part, our conventions in this section follow
those of \cite{nads2}.} and sets the stage for our subsequent calculations testing the 
CV and CA conjectures.

The fields of the JT model consist of a metric and a real-valued dilaton field, defined on a 1+1-dimensional
manifold $\mathcal{M}$ with timelike boundary $\partial\mathcal{M}$. By adopting a suitable field redefinition,
the action can be expressed in a simple form, with no derivatives acting on the dilaton,
\begin{eqnarray}
\mathcal{S}_{JT}&=& \frac{\varphi_0}{2} \Big(\int_\mathcal{M} d^2x \sqrt{-g}\,R
+2\int_{\partial\mathcal{M}} d\tau\,K\Big)
\nonumber \\
&\phantom{=}&\qquad+\frac12 \int_\mathcal{M} d^2x \sqrt{-g}\,\varphi\,\big( R+\frac{2}{L^2}\big)
+ \int_{\partial\mathcal{M}} d\tau\,\varphi\,\big(K-\frac{1}{L}\big)\,,
\label{JT_action}
\end{eqnarray}
where $L$ is a characteristic length scale of the model.

The boundary terms in the action involving the 
extrinsic curvature are needed to make the gravitational variational problem well defined and 
we have also included a boundary counterterm that renders the Euclidean on-shell action finite. 
The terms on the first line are independent of the dilaton field $\varphi$ and are instead 
multiplied by a constant $\varphi_0$. Their sum is proportional to a topological invariant, the 
Euler character of the 1+1-dimensional manifold, and thus they do not contribute 
to the equations of motion. 
They do, however, contribute to thermodynamic quantities such as the free energy 
and entropy of the black hole. 
As we shall see in Section~\ref{sec:JTmodel} below, this model, including the topological terms,  
arises naturally when one considers the spherical reduction of 3+1-dimensional Einstein-Maxwell 
theory around near-extremal charged black holes. In this context, the higher-dimensional black 
hole charge $Q$ determines the constant $\varphi_0$ in front of the topological terms
via the relation $\varphi_0=Q^2/2$. 
For now, we want to explore the JT theory on its own terms as a 1+1-dimensional model
and for this purpose we can simply assume that $\varphi_0\gg 1$ but leave it otherwise undetermined.

The boundary of the 1+1-dimensional spacetime is taken to be along a curve of constant dilaton field, 
$\varphi\big\vert_{\partial M}=\varphi_B$, and the integration variable $\tau$ in the boundary terms in 
the action is the proper time along this boundary curve. 
We can introduce a separation of scales in the JT model by imposing the condition $\varphi_B\ll \varphi_0$.
For the time being, we impose this by hand, but in the higher-dimensional context it amounts to 
restricting the range of the 1+1-dimensional effective description to lie well inside the throat region of a 
near-extremal Reissner-Nordstr\"om black hole. 

\subsection{AdS$_2$ geometry and JT black holes}

The action~\eqref{JT_action} gives rise to the following field equations, 
\begin{eqnarray}
0 &=& R + \frac{2}{L^2} \,,\label{dilatoneq} \\
0&=& \nabla_\alpha\nabla_\beta \varphi -g_{\alpha\beta}
\Big( \nabla^2\varphi-\frac{1}{L^2}\,\varphi\Big) \,. \label{einsteineq}
\end{eqnarray}
The variational equation for the dilaton field \eqref{dilatoneq} 
implies that the spacetime geometry is always locally AdS$_2$.  We find it convenient to
use global coordinates $(\nu,\sigma)$, where the AdS$_2$ metric takes the form
\be
ds^2 = \frac{L^2}{\sin^2\sigma}\,(-d\nu^2+d\sigma^2)\,.
\label{ads2metric}
\ee
The full range of the coordinates is $-\infty<\nu<\infty$ and $0<\sigma<\pi$, but 
when we take the dilaton field into consideration the coordinate range is restricted by the
boundary at $\varphi=\varphi_B$. 
This is best illustrated by considering a specific field configuration. 

\begin{figure}[htbp] 
   \centering
   \includegraphics[width=1.6in]{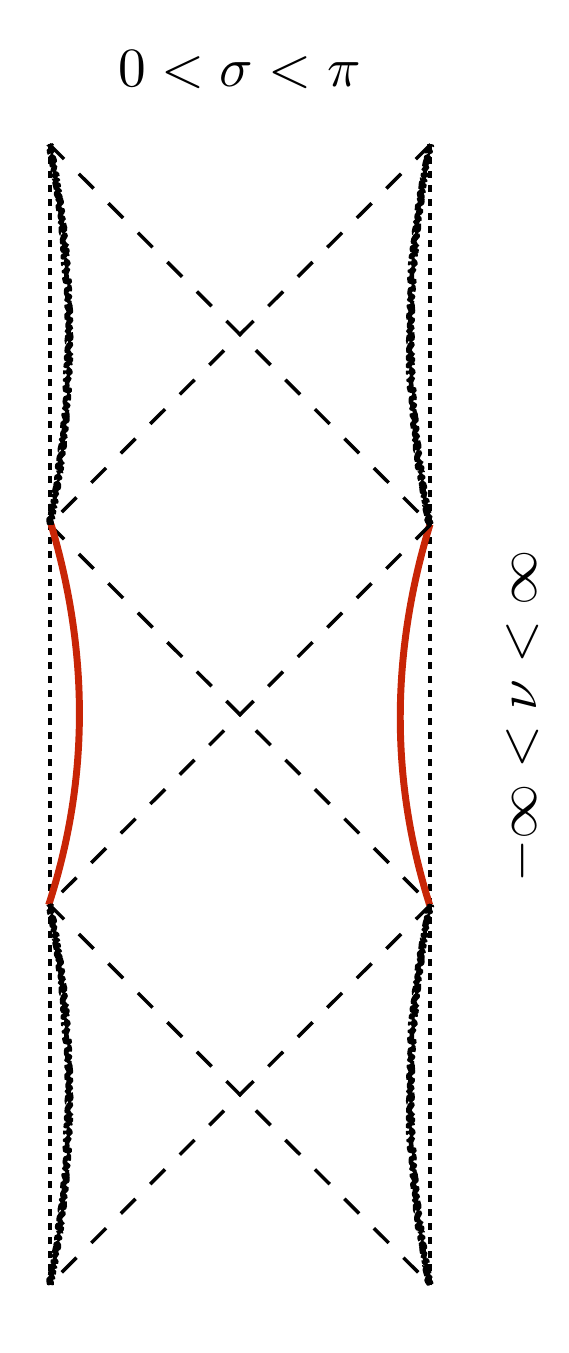} 
 \caption{Jackiw-Teitelboim black hole in global AdS$_2$ coordinates. 
 The {\color{red} red curves} indicate the outer boundary of the 1+1-dimensional black hole spacetime, while
 the solid black curves indicate singularities where $\Phi=0$. The dashed diagonal lines 
 show the location of the black hole horizon.}
   \label{fig:JTblackhole}
\end{figure}

A one-parameter family of solutions to the Einstein equations~\eqref{einsteineq} is given by
\be
\varphi(\nu,\sigma) = \varphi_H\,\frac{\cos\nu}{\sin\sigma} \,,
\label{dilaton_sol}
\ee
where the integration constant $\varphi_H>0$ is the value of $\varphi$ on the event horizon of the 
two-dimensional black hole, as we will see shortly.
The boundary at $\varphi=\varphi_B$ is located on a set of timelike curves,
\be
\sin\sigma = \varepsilon \cos\nu\,,
\label{boundarycurve}
\ee
with $\varepsilon=\varphi_H/\varphi_B$, that intersect the AdS$_2$ boundary at $\nu=\pm\frac{\pi}{2}$, 
as shown in Figure~\ref{fig:JTblackhole}.  
In the following we assume that $\varepsilon\ll 1$ and work to leading 
order in powers of $\varepsilon$. 

The dilaton field $\varphi$ in \eqref{dilaton_sol} is periodic in the global time $\nu$. It is
negative in the range $\frac{\pi}{2}<\nu<\frac{3\pi}{2}$ and 
becomes arbitrarily negative as $\sigma\rightarrow 0$ or $\sigma\rightarrow \pi$. 
This means that $\Phi\equiv\varphi_0+\varphi$ goes to zero along another set of timelike curves 
(also shown in Figure~~\ref{fig:JTblackhole}).  In the spherical reduction from 3+1-dimensions
discussed in Section~\ref{sec:JTmodel}, the field 
$\Phi$ is proportional to the area of the transverse two-sphere and with the higher-dimensional 
interpretation in mind it is natural to view the $\Phi=0$ curves as singularities.
The associated event horizon is along the diagonal 
lines $\nu=\pm (\sigma-\frac{\pi}{2})$ and it is easily checked that $\varphi=\varphi_H$ everywhere 
on the horizon. 
We note that the spherically reduced Reissner-Nordstr\"om solution, given in \eqref{2dBH} below, 
indeed has a timelike curvature singularity precisely where $\Phi$ vanishes. However, the small 
$\varphi$ truncation that leads to the field equations of the JT model breaks down long before $\varphi$ 
approaches $-Q^2/2$ so the $\Phi=0$ curve of the JT model is only a proxy for the true physical singularity.

\subsection{JT black hole thermodynamics}
\label{JTthermo}
The relationship between $\Phi$ and the area of the transverse two-sphere will be derived in Sec.~\ref{sec:JTmodel} and is given in \eqref{metric_ansatz}. The corresponding entropy of the 1+1-dimensional black hole is
\be
S=2\pi \Phi\big\vert_\textrm{Horizon}=2 \pi \varphi_0+ 2\pi \varphi_H\,.
\label{2d_entropy}
\ee
This entropy assignment can of course be made 
without giving it any higher-dimensional interpretation but in that case it would appear rather arbitrary.
The entropy of an extremal 3+1-dimensional Reissner-Nordstr\"om black hole is 
$S_0=\pi Q^2=2\pi\varphi_0$ so we see that $2\pi\varphi_H$ 
amounts to the added entropy of a near-extremal black hole with the same charge. 

The black hole character of the solution \eqref{dilaton_sol} is more immediately apparent when we
write it in a 1+1-dimensional version of Schwarzschild coordinates, 
\be
ds^2 = -\frac{r^2-r_H^2}{L^2}\,dt^2+\frac{L^2}{r^2-r_H^2}\,dr^2\,.
\label{s_coordinates}
\ee
\begin{figure}[htbp] 
   \centering
\includegraphics[width=1.8in]{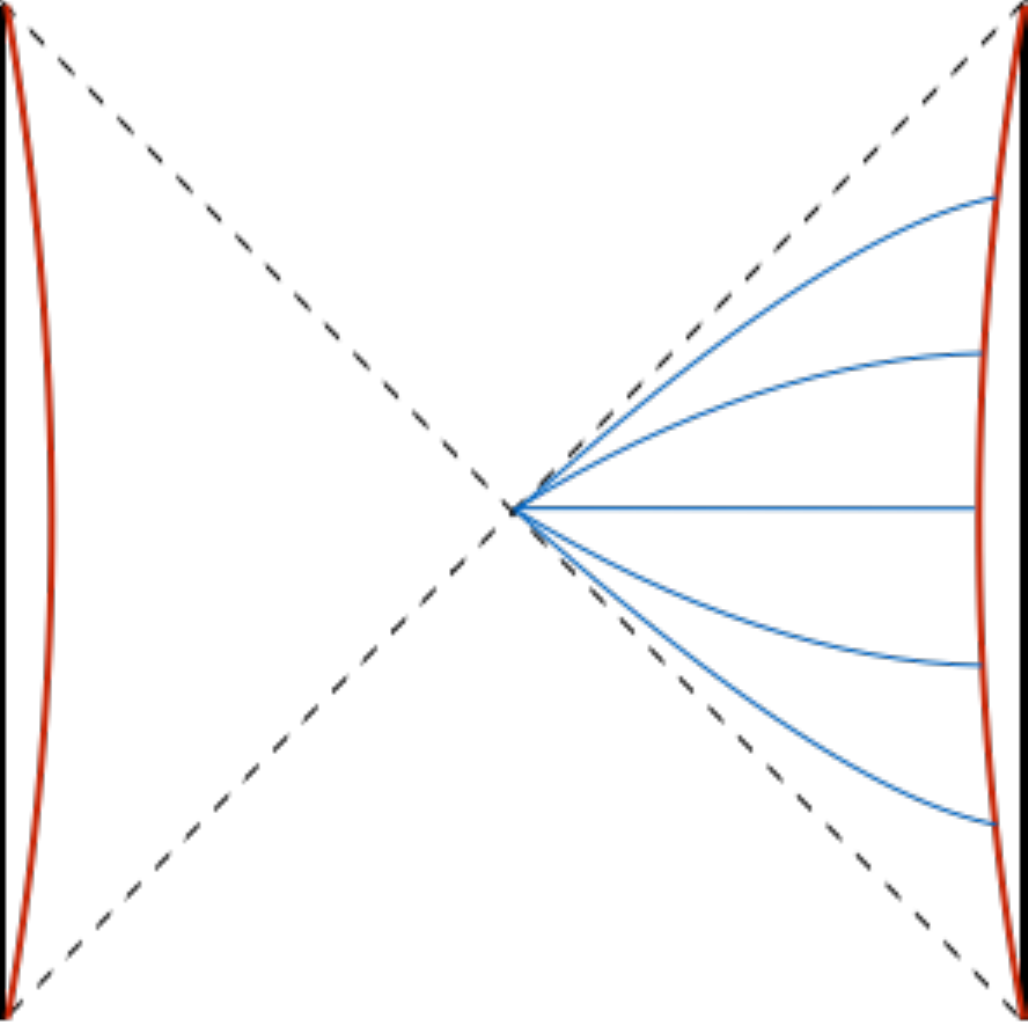}
 \caption{Jackiw-Teitelboim black hole: Curves of constant Schwarzschild time outside the event horizon
 are shown in {\color{blue} blue}.}
   \label{fig:s_time_slices}
\end{figure}
 In this coordinate system, there is a coordinate singularity at the event horizon 
at $r=r_H$ and the dilaton is linear in $r$,
\be
\varphi = \varphi_H\,\frac{r}{r_H}\,.
\label{lin_dil}
\ee
The requirement that the Euclidean continuation of the metric \eqref{s_coordinates} 
is smooth at $r=r_H$
yields the following Hawking temperature for the 1+1-dimensional black hole,
\be
T=\frac{r_H}{2\pi L^2}\,.
\label{2d_bh_temp}
\ee  
It is straightforward to work out the Kruskal extension of the Schwarzschild form of the
AdS$_2$ metric \eqref{s_coordinates} and find an explicit coordinate transformation 
relating $(\nu,\sigma)$ and $(t,r)$. In particular, the relationship between
the global time and Schwarzschild time along the boundary curve at $r=r_B$ is
\be
\tan{\Big(\frac{\nu}{2}+\frac{\pi}{4}\Big)}=e^{2\pi T t},
\label{timerelation}
\ee
up to $O(\varepsilon^2)$ correction terms that are small in the limit $r_B\gg r_H$. 
Slices of constant Schwarzschild time are shown in Figure~\ref{fig:s_time_slices}.

The on-shell Euclidean action can be obtained from the Euclidean version of 
\eqref{JT_action} including all the boundary terms. A straightforward calculation gives
\be
\mathcal{S}_E=-S+\beta E\,,
\label{2d_free_energy}
\ee
where
\begin{eqnarray}
S&=&2\pi \varphi_0+4\pi^2L^2\frac{\varphi_B}{r_B}\, T\,, \\
E&=& 2\pi^2L^2\frac{\varphi_B}{r_B}\, T^2 \,,
\end{eqnarray}
which has the form of a free energy of a near-extremal Reissner-Nordstr\"om black hole 
in a fixed charge ensemble (see Appendix~\ref{thermodynamics}).
By comparing with \eqref{RN_thermo} we identify 
$S$ as the 3+1-dimensional black hole entropy 
and $E$ as the added mass of the near-extremal black hole compared to the extremal mass. 
Furthermore, by equating the coefficients of $T$ and $T^2$ in $S$ and $E$ respectively, of the
1+1-dimensional black hole to the corresponding terms in \eqref{RN_thermo} one finds that 
$\varphi(r) = r/\ell$, where $\ell$ is the 3+1-dimensional Planck length.

\section{Complexity = Volume in the JT model}\label{sec:C=V} 

We are now in a position to test proposals for the holographic dual of quantum complexity
in the simplified setting of 1+1-dimensional dilaton gravity. We start with the complexity-volume (CV)
duality and then consider the complexity-action (CA) duality in Section~\ref{sec:CvsA}. 
The CV proposal \cite{Susskind:2014rva} states that the complexity of the quantum state dual 
to the black hole is proportional to the spatial volume of a maximal slice behind the horizon,
\be
\mathcal{C}\sim \frac{V}{G\,\ell_0}\,,
\label{CVproposal}
\ee
where $G$ is Newton's constant and $\ell_0$ is a characteristic length scale associated
with the black hole in question. In the case at hand, the `volume' of a maximal slice is 
simply the length $L_s$ of a spacelike geodesic and the characteristic length 
scale $\ell_0$ can be taken to be the AdS$_2$ scale $L$. There is, however, no notion of 
Newton's constant that is intrinsic to 1+1-dimensional gravity and we need to adapt the 
prescription accordingly. Our proposal is to include a factor of $\varphi_0$ in the proportionality 
factor between complexity and volume in the JT model, 
\be
\mathcal{C}\sim \frac{\varphi_0\,L_s}{L}\,.
\label{CV_JT}
\ee
This is motivated by the general expectation that the complexity should grow
at a rate that is proportional to the number of degrees of freedom of the dual 
quantum system. The number of degrees of freedom is in turn proportional to the 
black hole entropy and for our 1+1-dimensional black holes the 
entropy is dominated by the extremal entropy, $S_0=2\pi\varphi_0$. This is of course
also intimately related to the higher-dimensional origins of the Jackiw-Teitelboim model.
Under spherical reduction, each 1+1-dimensional event is accompanied
by a transverse sphere, whose area in Planck units is given by the dilaton field, and from
this point of view the 1+1-dimensional definition \eqref{CV_JT} is a special case of 
\eqref{CVproposal}.

\begin{figure}[htbp] 
   \centering
   \includegraphics[width=2.3in]{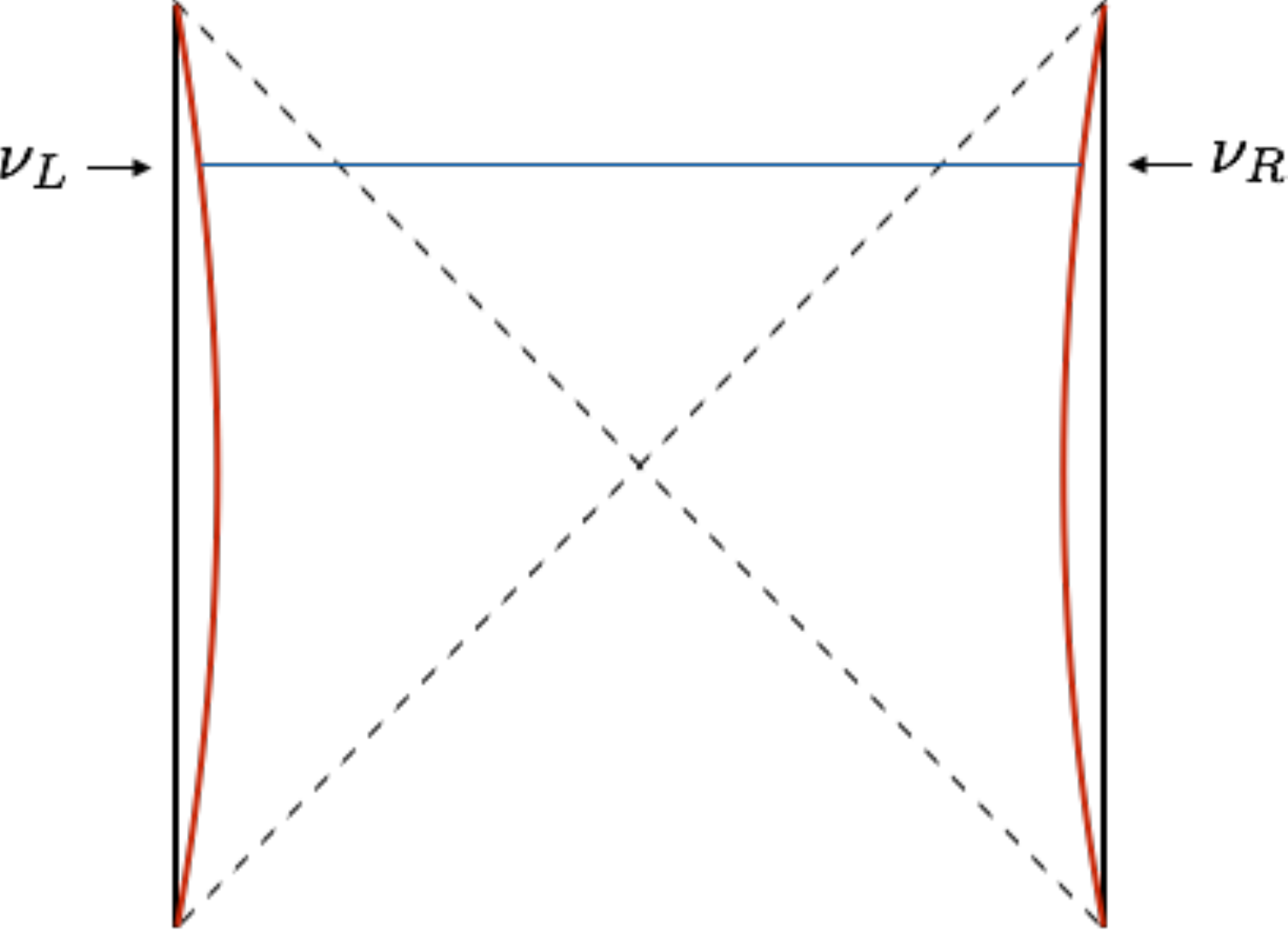} 
   \hspace{0.6in}
   \includegraphics[width=2.3in]{geodesic_1.pdf} 
 \caption{Minimal slices for $C=V$ calculation. The left panel shows a `horizontal' slice with 
 $\nu_L=\nu_R$. On the right, a more general slice with independent values of $\nu_L$, $\nu_R$.}
   \label{fig:C=V_slices}
\end{figure}

Let us first consider a maximal slice of the form shown in the left panel of Figure~\ref{fig:C=V_slices},
where the spacelike geodesic connects boundary points that have the same value of global
time, $\nu_L=\nu_R\equiv\nu_0$. The complexity is proportional to the proper length that lies inside
the horizon,
\begin{eqnarray}
L_s&=&L\int_{\frac{\pi}{2}-\nu_0}^{\frac{\pi}{2}+\nu_0} \frac{d\sigma}{\sin{\sigma}} \nonumber \\
&=& 2L\log\left(\tan\left(\frac{\nu_0}{2}+\frac{\pi}{4}\right)\right) \nonumber \\
&\approx& 4\pi L T t_0  \,,
\end{eqnarray} 
where in the final step we have used \eqref{timerelation} to convert from global time to Schwarzschild time.
It then immediately follows from \eqref{CV_JT} that complexity-as-volume grows 
linearly with Schwarzschild time in the JT model,
\be
C\sim \phi_0 T t \,,
\ee
and the rate of growth is proportional to the black hole temperature $dC/dt \sim S T$.  

More generally, we can consider a spacelike geodesic that connects left- and right-hand boundary points
at different global time, $\nu_L\neq\nu_R$, as shown in the right hand panel of Figure~\ref{fig:C=V_slices}.
A geodesic satisfying these boundary conditions (up to $O(\varepsilon^2)$ corrections) is given by the 
following curve in the $(\nu,\sigma)$ plane,
\be
\sin{(\nu-\nu_+)}=\sin{\nu_-}\,\cos{\sigma} \,,
\label{geodesiccurve}
\ee
where $\nu_\pm =\frac12 (\nu_L\pm\nu_R)$. The geodesic meets the horizon $\sigma =\frac{\pi}{2}\pm \nu$ at 
two intersection points at $\sigma=\sigma_1$ and  $\sigma=\sigma_2$ as indicated in Figure~\ref{fig:C=V_slices}. 
The locations of the intersection points are easily obtained from the geodesic curve and are given by the relations
\be
\tan{\sigma_1}=\frac{(\cos{\nu_+}-\sin{\nu_-})}{\sin{\nu_+}}\>;\qquad
\tan{\sigma_2}=-\frac{(\cos{\nu_+}+\sin{\nu_-})}{\sin{\nu_+}}\,.
\label{sigmarelations}
\ee
The geodesic length inside the horizon is then given by the integral
\be
L_s=L\int_{\sigma_1}^{\sigma_2} \frac{d\sigma}{\sin{\sigma}} \sqrt{1-\left(\frac{d\nu}{d\sigma}\right)^2},
\ee
with $\nu(\sigma)$ obtained from \eqref{geodesiccurve}. This yields a relatively simple closed 
form expression involving $\sigma_1$ and $\sigma_2$,
\be
L_s= \frac{L}{2}\left[ \log\left(\frac{1-\cos\sigma}{1+\cos\sigma}\right)
+\log\left(\frac{1+\sin^2\nu_-\cos\sigma+\cos\nu_-\sqrt{1-\sin^2\nu_-\cos^2\sigma}}
{1-\sin^2\nu_-\cos\sigma+\cos\nu_-\sqrt{1-\sin^2\nu_-\cos^2\sigma}}\right)\right]_{\sigma_1}^{\sigma_2},
\ee
which can in turn be expressed in terms of $\nu_L$ and $\nu_R$ via \eqref{sigmarelations}. 

As a side note, in the Schwarzian regime \cite{nads2} this formula simplifies simplifies when the usual Schwarzian variables $\nu_L(u)$ and $\nu_R(u)$ are used, where $u$ is the proper time along the boundary trajectories:
\be 
L_s = -\log  \left[ {\nu_L'(u) \nu_R'(u)\over \cos^2({\nu_L - \nu_R \over 2 })} \right] + \mathcal{O}(\epsilon^2).
\ee
In this expression, we have dropped the constant IR regulator $\sim \log \epsilon^2$. We see explicitly that the volume is invariant under $SL(2,R)$ transformations. In fact, we could have guessed the form of this expression based on $SL(2,R)$ symmetry and the fact that the length is only a function of $\nu$ and its first derivative. 

We are mainly interested in the late-time limit on one boundary, {\it i.e.} $\nu_R\rightarrow \frac{\pi}{2}$ while 
keeping $\nu_L$ fixed. From Figure~\ref{fig:C=V_slices} it is immediately apparent that $\sigma_2\rightarrow\pi$
in this limit, and the geodesic length inside the horizon is dominated by a single term,
\begin{eqnarray}
L_0&=&- \frac{L}{2}\log\left(1+\cos\sigma_2\right)+\ldots \nonumber\\
&=&-L \log(\cos\nu_R)+\ldots \nonumber\\
&\approx& \> 2\pi LT t_R +\ldots\,.
\end{eqnarray}
The ``$\ldots$'' refers to terms that remain finite as $t_R\rightarrow\infty$ while keeping $t_L$ fixed.
We again find that the complexity-as-volume grows linearly with time at late times, but now at half 
the rate we found previously when both $t_L$ and $t_R$ progressed towards late times. 
We conclude that the CV conjecture makes similar predictions for 1+1-dimensional black holes as it 
does in higher dimensions, provided we `by hand' multiply the geodesic length inside the horizon by 
a factor of the black hole entropy. 

One comment on the volume calculation is that the volume of the wormhole is determined entirely by the 
motion of the boundary, which in turn is governed by the Schwarzian effective action. In any quantum 
mechanical model such as the SYK model, with a spontaneously and explicitly broken time-reparameterization 
symmetry, it is expected that the Schwarzian action will dominate at low energies due to the $N/\beta J$ enhancement. 
If the CV conjecture is correct, any quantum mechanical model with such a nearly conformal symmetry should 
exhibit a complexity that matches the above calculations in the appropriate low-energy regime. 
In other words, we do {\it not} expect that an improved understanding of the bulk dual of SYK will 
change the calculation of the volume, as long as we are not interested in exponentially late times or 
energies outside of the low-energy regime $\beta J \lesssim 1$.
The CV conjecture thus suggests that the complexity in an NCFT$_1$ is universal in 
the above sense. It would be interesting to understand this better from the boundary perspective.

\section{Complexity $\neq$ Action in the JT model}\label{sec:CvsA} 

\begin{figure}[htbp] 
   \centering
   \includegraphics[width=2.5in]{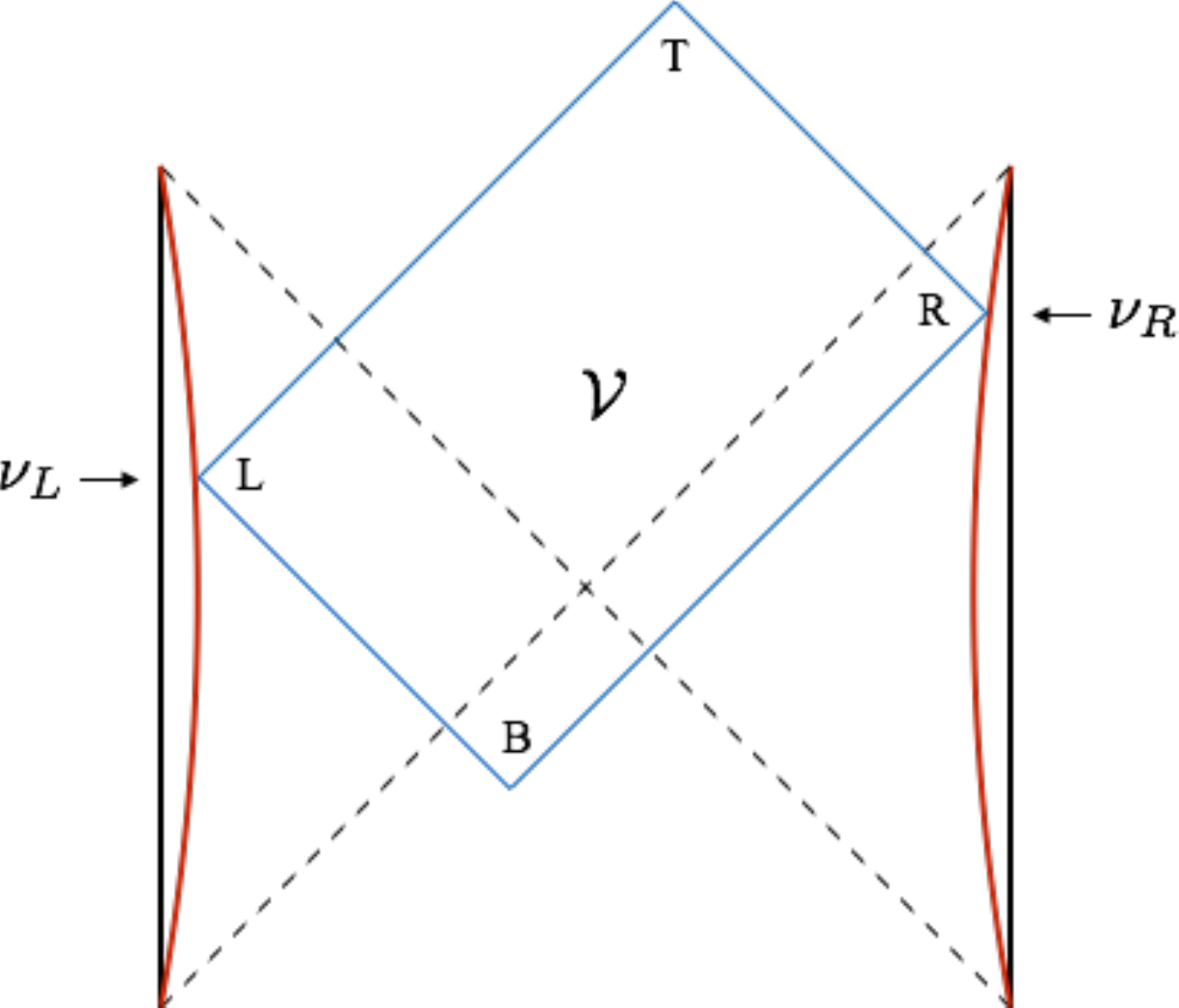} 
 \caption{Wheeler-DeWitt patch for CA calculation.}
   \label{fig:WDW_patch}
\end{figure}

In this section we carry out a test of CA duality by evaluating the classical action of the
JT model on a Wheeler-DeWitt (WDW) patch anchored on boundary points at $\nu=\nu_L$ and
$\nu=\nu_R$ as shown in Figure~\ref{fig:WDW_patch}. An immediate problem that arises
is that the boundary of the WDW patch includes segments of null curves but the boundary
terms in the action \eqref{JT_action} involving the extrinsic curvature are only well defined if the
boundary is spacelike or timelike.
Another problem is that the boundary of the WDW patch fails to be everywhere smooth. 
The null segments that make up the boundary meet at sharp corners and the corner points
will contribute to the bulk action even if they are of co-dimension two. Both these issues, 
{\it i.e.} the degeneracy of extrinsic curvature terms on null boundary segments and the corner 
contributions, can be addressed by applying a limiting procedure involving a family of timelike 
curves that approach the null boundary curves as a parameter is varied. 
In the following, we will adopt the prescription of Lehner~{\it et al.\/}\ \cite{Lehner:2016vdi} for 
calculating the action on a WDW patch, which  
can be applied directly to intersecting null boundary segments and is known to be equivalent to
the original results of \cite{Brown:2015bva,Brown:2015lvg} for various black hole geometries. 

The boundary of the WDW patch $\mathcal V$ in Figure~\ref{fig:WDW_patch} consists of 
four null segments that intersect pairwise at four corner points labeled by $i\in\{L,T,R,B\}$. 
According to the prescription of \cite{Lehner:2016vdi}, the contribution from a null boundary  
is given by an integral whose value depends on how the boundary curve is parametrised. 
The choice of boundary parametrisation also affects the value of the corner contributions but
when all the boundary and corner terms are added up the total is independent of 
parametrisation. In particular, if we choose affine parametrisation, so that each boundary 
segment is a null geodesic connecting two corner points, then the boundary integrals 
vanish and the action on the WDW patch consists of bulk and corner contributions only,
\be
S_\textrm{WDW}=\frac{\varphi_0}{2}\int_\mathcal{V}d^2x \sqrt{-g}R+\varphi_0\sum_i a_i
+\frac{1}{2}\int_\mathcal{V}d^2x \sqrt{-g}\,\varphi\Big(R+\frac{2}{L^2}\big)+\sum_i \varphi_i\, a_i \, ,
\label{wdw_action}
\ee
where $\varphi_i$ is the value of the dilaton field at the $i$-th corner point and 
$a_i = \pm \log\big\vert \frac12 (k\cdot k')\big\vert$,
with $k$ and $k'$ future directed tangent vectors of the null segments that intersect at
the corner point in question, subject to a certain normalisation condition at the JT boundary
that ensures the parametrisation-independence of the overall result.
The sign of the corner contribution $a_i$ associated to a given corner point is determined 
as follows \cite{Lehner:2016vdi}: 
Consider either of the two null boundary segments that intersect at the corner in question. 
The sign is positive if the bulk region $\mathcal{V}$ lies to the future (past) of the segment and the 
corner point is at the past (future) end of the segment. Otherwise the sign is negative. 
This translates into a positive sign for the top and bottom corners in Figure~\ref{fig:WDW_patch}
and a negative sign for the left and right corners. 
This procedure gives an action that is invariant under the $SL(2,R)$ gauge symmetry  
which acts on the boundaries by an AdS$_2$ isometry \cite{nads2}. These $SL(2,R)$ transformations 
can be viewed as coordinate transformations, and our procedure for regulating the action 
(or the volume) is manifestly coordinate-invariant. This is of course a desirable feature of any 
regulator, since complexity is a gauge-invariant quantity.

The bulk terms in the action on the WDW patch \eqref{wdw_action} are easily evaluated.
The curvature scalar is given by $R=-2/L^2$ everywhere so the bulk JT term is manifestly zero.
Therefore the only bulk contribution comes from the topological term and the integral 
over the WDW patch gives
\begin{eqnarray}
S_\textrm{bulk}&=&-\varphi_0\int_\mathcal{V} d\nu d\sigma\,\frac{1}{\sin^2\sigma} \nonumber\\
&=& 2\,\varphi_0\big(\log \left(\sin\sigma_L\right)+\log \left(\sin\sigma_R\right)
-\log \left(\sin\sigma_T\right)-\log \left(\sin\sigma_B\big)\right).
\label{topological_bulk}
\end{eqnarray}

The corner terms are also easily evaluated once we settle on a choice of 
parametrization for each null boundary segment $(\nu(\lambda),\sigma(\lambda))$.
In general the parametrisation
can be arbitrary but for reasons explained above we will assume $\lambda$ is an 
affine parameter. It is easy to check that a given boundary segment is indeed a 
geodesic if the components of its (future directed) tangent vector have the form
\be
\frac{d\nu}{d\lambda}=\pm \frac{d\sigma}{d\lambda}= A\,\sin^2\sigma(\lambda)\,,
\label{tangent_vector}
\ee
for some normalisation constant $A>0$. Different values of $A$ correspond to the usual freedom
to rescale the affine parameter. The prescription of \cite{Lehner:2016vdi} is to impose a normalisation
condition on the tangent vector involving its inner product with the timelike Killing vector 
that generates Schwarzschild time translations at the JT boundary,
\be
\big(\frac{d\nu}{d\lambda},\frac{d\sigma}{d\lambda}\big)\cdot 
\big(\frac{d\nu}{dt},\frac{d\sigma}{dt}\big) = -c\, ,
\label{k_normalization}
\ee
where $c>0$ is a constant. 
The inner product is to be evaluated at the corner point where the boundary segment under
consideration meets the JT boundary.
It is important that $c$ is time independent, {\it i.e.} that the same value of $c$ is chosen 
when normalising boundary tangent vectors for WDW patches anchored at different boundary times,
but $c$ is otherwise a free parameter. As it turns out, the formulas below simplify 
for $c=L$ and we will choose this value throughout. 
The normalisation constant in \eqref{tangent_vector} is then given by
\be
A=\Big(L\,\frac{d\nu_L}{dt_L}\Big)^{-1} \qquad\textrm{or} \qquad
A=\Big(L\,\frac{d\nu_R}{dt_R}\Big)^{-1} ,
\ee
depending on whether the boundary segment in question intersects the left or right hand boundary
in Figure~\ref{fig:WDW_patch} and the corner factors $a_i$ in \eqref{wdw_action} reduce to 
\begin{eqnarray}
a_L&=&-2\log \left(\sin\sigma_L\right)+2\log\Big(\frac{d\nu_L}{dt_L}\Big), \nonumber \\
a_R&=&-2\log \left(\sin\sigma_R\right)+2\log\Big(\frac{d\nu_R}{dt_R}\Big), \nonumber \\
a_T&=& 2\log \left(\sin\sigma_T\right)-\log\Big(\frac{d\nu_L}{dt_L}\,\frac{d\nu_R}{dt_R}\Big), \nonumber \\
a_B&=& 2\log \left(\sin\sigma_B\right)-\log\Big(\frac{d\nu_L}{dt_L}\,\frac{d\nu_R}{dt_R}\Big). 
\end{eqnarray}
The total contribution to the action from the corners of the WDW patch is 
\be
S_\textrm{corners}=\varphi_0\sum_i a_i +\sum_i \varphi_i \, a_i \,.
\label{cornerterms}
\ee
The first sum comes from the topological part of the action \eqref{JT_action} while the second
comes from the dynamical part. 

The sum over the topological corner terms evaluates to
\be
\varphi_0\sum_i a_i 
= 2\,\varphi_0\big(-\log \left(\sin\sigma_L\right)-\log \left(\sin\sigma_R\right)
+\log \left(\sin\sigma_T\right)+\log \left(\sin\sigma_B\big)\right),
\ee
which precisely cancels the topological bulk contribution \eqref{topological_bulk}.

This leaves us with the very simple result that the sole contribution to the WDW patch action 
comes from the sum over non-topological corner terms in \eqref{cornerterms}. Let us first consider the
left and right corners, which are located on the JT boundary, where the dilaton takes the value
$\varphi=\varphi_B$ independent of the boundary time. It turns out, the corner factors $a_{L,R}$ are 
also time-independent at leading order in $\varepsilon$ and thus the overall contribution from the
left and right hand corners to the WDW action is constant in time.\footnote{This is in line with the general 
expectation that the contribution to the WDW patch action from outside the horizon 
does not depend on the boundary time at which the WDW patch is anchored. } 
To see this, we first use \eqref{timerelation} to evaluate the Jacobian factor between global 
time and Schwarzschild time,
\be
\frac{d\nu_{L,R}}{dt_{L,R}}=2\pi T\,\cos{\nu_{L,R}} \,,
\ee
and then use that $\sin\sigma_{L,R}=\varepsilon\cos\nu_{L,R}$ on the JT boundary to find
\be
a_{L,R}=2\log\Big(\frac{4\pi T}{\varepsilon}\Big)\,.
\ee

Finally, we consider the top and bottom corners. 
Working to leading order in $\varepsilon$ one finds that 
\be
\sigma_T = \frac{\pi}{2}-\frac{\nu_l}{2}+\frac{\nu_R}{2}\,,\qquad\qquad
\sigma_B = \frac{\pi}{2}+\frac{\nu_l}{2}-\frac{\nu_R}{2}\,,
\ee
and then a short calculation gives
\be 
a_T=a_B=2\log\cos\Big(\frac{\nu_L-\nu_R}{2}\Big)
-\log\Big(\frac{d\nu_L}{dt_L}\,\frac{d\nu_R}{dt_R}\Big)  \,.
\ee
It follows that both the top and bottom corner factors grow linearly with time,
\be 
a_{T,B}\approx 2\pi LT t_{L,R} +\ldots\,,
\ee
as $t_{L,R}\rightarrow\infty$. However, their combined contribution to the WDW patch 
action cancels at late times due to the dilaton prefactors in \eqref{cornerterms}. To see this, 
consider the WDW patch in Figure~\ref{fig:WDW_patch} and let either $t_L$ or $t_R$
approach infinity, {\it i.e.} let $\nu_L\rightarrow \frac{\pi}{2}$ or $\nu_R\rightarrow \frac{\pi}{2}$. 
Either way, the top corner of the WDW patch will approach a point on the inner horizon,
where $\phi=-\phi_H$, while the bottom corner approaches a point on the outer horizon,
where $\phi=+\phi_H$. The leading linear growth thus cancels between the top and 
bottom corners. 

Bringing everything together we reach the conclusion that the action on a WDW patch in 
Jackiw-Teitelboim gravity does not grow linearly with time at late boundary times but instead 
approaches a constant value. This is a surprising result in view of the conjectured duality 
between low-energy sectors of JT gravity and the SYK model. In the latter, quantum complexity 
is expected to grow linearly for a very long time. The JT model prediction for complexity is also
at odds with \eqref{eq:chargedincrement}, which gives the rate of growth of the action on a 
WDW patch of a Reissner-Nordstr\"om black hole in 3+1-dimensional Einstein-Maxwell theory.
The discrepancy can be resolved but only after a careful re-examination of how JT gravity is 
obtained from the higher-dimensional theory, to which we now turn. 

\section{JT gravity from dimensional reduction}
\label{sec:JTmodel}

Consider the Reissner-Nordstr\"om black holes described in Sec.~\ref{sec:RN_C=A}. 
In the near-extremal limit, the throat has approximately constant width and is very long, 
and supports a low-energy sector of radial excitations that is governed by an effective 
two-dimensional theory. This effective theory is JT gravity \cite{NavarroSalas:1999up, Nayak:2018qej}. To derive the action for JT gravity from the 3+1-dimensional Einstein-Maxwell theory, Navarro-Salas \& Navarro 
integrated out the transverse directions \cite{NavarroSalas:1999up}. 
We review that reduction here. 

The first step is to adopt an ansatz for a spherically symmetric metric,
\begin{equation}
ds^2=\frac{1}{\sqrt{2\Phi}} g_{\alpha\beta}\, dx^\alpha dx^\beta+2\ell^2\Phi\, d\Omega^2\,,
\label{metric_ansatz}
\end{equation}
and insert it into into the 3+1-dimensional action \eqref{4d_action}.
Here $g_{\alpha\beta}(x^0,x^1)$ is a 1+1-dimensional metric 
and the dilaton $\Phi(x^0,x^1)$ is a scalar field that describes how the area of the transverse
two-sphere depends on time and radial position. The resulting action is
\begin{equation}
\mathcal{S}_{2d}=\frac12 \int d^2x \sqrt{-g}\big(\Phi R+\frac{1}{\ell^2}(2\Phi)^{-\frac12}
-\frac{\ell^2}{2}(2\Phi)^{\frac32}F_{\alpha\beta}F^{\alpha\beta}\big) 
+\int dy^0 \sqrt{-\gamma_{00}}\big(\Phi K-\frac{1}{\ell}(2\Phi)^{\frac14}\big),
\label{2d_action}
\end{equation}
with the boundary terms evaluated along a timelike boundary with induced metric $\gamma_{00}$. 
The two-dimensional field strength $F_{\alpha\beta}$ is inherited unchanged from the 
3+1-dimensional theory but the contraction in the $F^2$ term in the action is now with the 
two-dimensional metric. The $\Phi$-dependent prefactor in front of $g_{\alpha\beta}$ in 
\eqref{metric_ansatz} implements a Weyl transformation on the two-dimensional metric that 
eliminates derivative terms involving $\Phi$ from \eqref{2d_action}.
Under spherical reduction, the extrinsic curvature term in the original 3+1-dimensional action 
\eqref{4d_action} gives rise to the boundary term containing the one-dimensional 
extrinsic curvature in \eqref{2d_action} 
and also a term involving the normal derivative of the dilaton field on the boundary. This latter term 
cancels against a total derivative term involving the dilaton that comes from the 3+1-dimensional 
Ricci scalar. The last term in \eqref{2d_action} 
comes from the spherical reduction of the $K_0$ regulator term in the original action.
Finally, if the electromagnetic boundary term \eqref{maxwellboundaryterm} is included in the 
3+1-dimensional action, then the 1+1-dimensional action will include its spherical reduction,
\begin{equation}
\mathcal{S}_{b,2d}^\textrm{em}
=\ell^2 \int dy^0 \sqrt{-\gamma_{00}}\,(2\Phi)^{\frac32}\,\hat{n}_\alpha\, F^{\alpha\beta}A_\beta \,,
\label{maxwell_1d}
\end{equation}
as an additional boundary term.

The field equations of the 1+1-dimensional theory are,
\begin{eqnarray}
0&=& \nabla_\alpha\left(\Phi^{3/2}F^{\alpha\beta}\right), \label{2dMaxwell} \\
0&=& R-\frac{1}{\ell^2}(2\Phi)^{-3/2}-\frac32 \ell^2 (2\Phi)^{1/2}F^2\,, \label{2dDilaton} \\
0&=& \nabla_\alpha\nabla_\beta \Phi -g_{\alpha\beta}
\Big( \nabla^2\Phi-\frac{1}{2 \ell^2}(2\Phi)^{-1/2} \Big)
+\ell^2 (2\Phi)^{3/2}\Big(F_{\alpha\gamma}F_\beta^{\phantom{\beta}\gamma}
-\frac14 g_{\alpha\beta}F^2\Big).
\label{2dEinstein} 
\end{eqnarray}
The Maxwell equation determines the electromagnetic field strength in terms of the dilaton,
\begin{equation}
F_{\alpha\beta}=\frac{Q}{\ell^2}(2\Phi)^{-3/2} \, \varepsilon_{\alpha\beta}\,,
\label{Fstrength}
\end{equation}
where $\varepsilon_{\alpha\beta}$ is the two-dimensional Levi-Civita tensor,\footnote{With the convention
$\varepsilon_{01}=+\sqrt{-g}$.} and this can be used to eliminate $F_{\alpha\beta}$ from the remaining field equations, 
\begin{eqnarray}
0&=& R-\frac{1}{\ell^2}(2\Phi)^{-3/2}+\frac{3Q^2}{\ell^2} (2\Phi)^{-5/2}\,, \label{2dDilaton2} \\
0&=& \nabla_\alpha\nabla_\beta \Phi -g_{\alpha\beta}
\Big( \nabla^2\Phi-\frac{1}{2 \ell^2}(2\Phi)^{-1/2}
+\frac{Q^2}{2\ell^2} (2\Phi)^{-3/2} \Big).
\label{2dEinstein2} 
\end{eqnarray}
We note that these equations are satisfied by the dimensional reduction of the 
Reissner-Nordstr\"om solution,
\eqref{RNsolution}, 
\begin{equation}
ds^2= -\Big(\sqrt{\frac{2x}{\ell}}-2\ell M+Q^2\sqrt{\frac{\ell}{2x}}\Big) dt^2 
+ \frac{dx^2}{\Big(\sqrt{\frac{2x}{\ell}}-2\ell M+Q^2\sqrt{\frac{\ell}{2x}}\Big)} \,,
\label{2dBH} 
\end{equation}
and a linear dilaton field $\Phi(x)= x/\ell$. The results of Appendix~\ref{thermodynamics} on charged 
black hole thermodynamics can be reproduced from the 1+1-dimensional Euclidean on-shell action, 
evaluated on this solution. In particular, the presence or absence of the Euclidean counterpart to the
spherically reduced electromagnetic boundary term \eqref{maxwell_1d} determines whether the
ensemble is at fixed chemical potential or fixed charge.  

In the following, we will mainly be interested in near-extremal black holes. More specifically, we want
to study the near horizon physics of a near-extremal black hole. For this purpose, we expand the 
dilaton field around its value at the horizon of an extremal black hole,
\begin{equation}
\Phi = \frac{Q^2}{2}+\varphi\,,
\label{near_horizon}
\end{equation}
and work order by order in $\varphi/Q^2$. At leading order, the field equations 
\eqref{2dDilaton2} and \eqref{2dEinstein2} reduce to,
\begin{eqnarray}
0 &=& R + \frac{2}{Q^3} \,,\label{adscurvature} \\
0&=& \nabla_\alpha\nabla_\beta \varphi -g_{\alpha\beta}
\Big( \nabla^2\varphi-\frac{1}{Q^3}\,\varphi\Big) \,, \label{varphi_eq}
\end{eqnarray}
which are precisely the field equations \eqref{dilatoneq} and \eqref{einsteineq} of the 
JT model with $L\equiv Q^{3/2} \ell$. 
It immediately follows that in the near-horizon region
the 1+1-dimensional radial geometry is that of AdS$_2$, with a characteristic 
length scale $L$ that is parametrically large compared to the 3+1-dimensional Planck length 
when $Q\gg 1$. This is the long throat of the near-extremal Reissner-Nordstr\"om black hole
referred to in Section~\ref{sec:RN_C=A}. 

We arrived at the reduced set of field equations by using the 1+1-dimensional Maxwell equations
to eliminate $F_{\alpha\beta}$ and it is natural to ask if the Jackiw-Teitelboim action \eqref{JT_action}
can similarly be obtained by integrating out the gauge field from the spherically reduced action
and considering the near-horizon limit. The answer is yes but with a somewhat subtle twist.
The most naive approach, where one simply inserts the solution \eqref{Fstrength} for 
$F_{\alpha\beta}$ into the full 1+1-dimensional action \eqref{2d_action} does not work.
This naive procedure does lead to a dilaton gravity theory but one where the term in the effective 
potential for the dilaton that comes from the gauge field has the wrong sign to reproduce the
Jackiw-Teitelboim theory in the near-horizon limit. The problem can be traced to the fact that
the gauge field we are integrating out is an electric field and we are replacing its kinetic energy
by an effective potential for the dilaton. In fact, this kind of sign flip occurs any time a dynamical 
variable carrying kinetic energy is integrated out in favor of a potential energy term. 

We illustrate this effect in 
Appendix~\ref{app:central_pot} using the familiar example of a non-relativistic
particle moving in a central potential. The analysis of particle orbits is facilitated by introducing
an effective potential for radial motion with a centrifugal term involving the conserved angular
momentum. This is usually done at the level of the equations of motion but if one instead attempts 
to integrate out the angular variable at the level of the Lagrangian before deriving the radial 
equation then an analogous sign issue arises. 
The remedy, both for motion in a central potential and in the case at hand, is to include 
appropriate boundary terms for the kinetic variable in the original action. Adding a boundary
term involving the gauge field does not change its dynamical equations, {\it i.e.} the Maxwell
equations are not affected, but a boundary term will in general contribute to the effective dilaton 
potential that results from integrating out the gauge field. 

As it turns out, we have already introduced a boundary term \eqref{maxwell_1d} that has the 
desired effect. To see this, we can use the divergence theorem to rewrite the boundary term as a 
1+1-dimensional bulk term involving a total derivative, apply the chain rule, and then use the 
Maxwell equation \eqref{2dMaxwell} to simplify the result,
\begin{eqnarray}
\mathcal{S}_{b,2d}^\textrm{em}
&=&\ell^2 \int d^2x \sqrt{-g}\, \nabla_\alpha \big((2\Phi)^{\frac32}\, F^{\alpha\beta}A_\beta\big) 
\nonumber \\
&=&\frac{\ell^2}{2} \int d^2x \sqrt{-g} (2\Phi)^{\frac32}\, F^{\alpha\beta}F_{\alpha\beta} \,.
\label{bulkform}
\end{eqnarray}
This has the same form as the electromagnetic bulk term in the 1+1-dimensional action
that we obtained from spherical reduction but has a coefficient in front that is twice as large 
and of opposite sign. This is precisely what is needed to reverse the sign of the 
electromagnetic contribution to the dilaton effective potential when we insert the solution
\eqref{Fstrength} for the Maxwell field into the action. The resulting bulk effective action is 
\begin{equation}
\mathcal{S}_\textrm{bulk}=\frac12 \int d^2x \sqrt{-g}\Big(\Phi R+\frac{1}{\ell^2}(2\Phi)^{-\frac12}
-\frac{Q^2}{\ell^2}(2\Phi)^{-\frac32}\Big).
\end{equation}
To capture the near-horizon physics of a near-extremal black hole
we write the dilaton as in \eqref{near_horizon} and work order by order in $\varphi$, 
\begin{equation}
\mathcal{S}_\textrm{bulk}= \frac{Q^2}{4} \int d^2x \sqrt{-g}\,R
+\frac12 \int d^2x \sqrt{-g}\,\varphi\,\Big( R+\frac{2}{L^2}\Big)+\ldots \,,
\label{2d_dilaton_action}
\end{equation}
This agrees precisely with the bulk terms in the Jackiw-Teitelboim action \eqref{JT_action}
if we make the identification $\varphi_0=\frac{Q^2}{2}$. The $\ldots$ denotes terms 
that are suppressed in the near-horizon region where $\varphi\ll Q^2$.
Further away from the horizon the additional terms are no longer small and 
this simple truncation does not apply. We can ensure that we stay inside the 
region of interest by introducing a boundary inside the AdS$_2$ throat region,
where \eqref{2d_dilaton_action} remains valid, and supplementing the bulk terms
by appropriate boundary terms. The boundary terms in \eqref{2d_action} are 
evaluated in the asymptotic region far outside the AdS$_2$ throat so we have to
to look elsewhere for a good definition of the boundary terms inside the throat.
A Gibbons-Hawking boundary term involving the extrinsic curvature of the 
boundary curve is needed in order to have a well posed variational problem 
for the 1+1-dimensional metric. The further requirement that the Euclidean 
on-shell action give a finite free energy is satisfied by including an additional 
boundary term. The final form of the 1+1-dimensional action of the near-horizon 
dilaton gravity theory, including the boundary terms, is then precisely the JT action 
\eqref{JT_action}. The boundary is placed along a curve of constant dilaton,
$\varphi\big\vert_{\partial M}=\varphi_B$. The requirement that the boundary be
inside the near-extremal AdS$_2\times S^2$ region translates into $\varphi_B \ll Q^2$.

\section{Complexity = Action restored in the JT model} \label{sec:theresolution}

We now return to our test of CA duality.
In Section~\ref{sec:CvsA} we evaluated the JT action on a 1+1-dimensional WDW patch
and found that it does not exhibit the expected linear growth at late times but instead 
approaches a constant value. This discrepancy can be traced to the procedure by 
which the JT model is obtained from higher-dimensional Einstein-Maxwell theory via 
dimensional reduction. The particular step we have in mind is where an electromagnetic
boundary term had to be introduced in order to get the correct sign for the dilaton potential 
when integrating out the 1+1-dimensional Maxwell field. The boundary term does not
affect the field equations of the theory but it does change the value of the action itself
and a choice has to be made whether to include it when calculating the action of a 
WDW patch. Our calculation in Section~\ref{sec:CvsA} is based on the JT action
\eqref{JT_action} and thus includes the boundary term in question. The calculation 
can easily be repeated without the boundary term included. Rather than starting 
from scratch, we can simply evaluate the contribution from the electromagnetic 
boundary term on the same WDW patch and subtract it from our previous answer
for the JT model. The most convenient way to proceed is to work with the equivalent
bulk form \eqref{bulkform} and evaluate it on-shell using the AdS$_2$ metric
\eqref{ads2metric} with $L^2=Q^3\ell^2$ and the identification $\varphi_0=Q^2/2$,
\begin{eqnarray}
\mathcal{S}_\textrm{WDW}&=&\mathcal{S}_\textrm{WDW}^\textrm{JT}
+\frac{Q^2}{\ell^2}\int_\mathcal{V}d^2x \sqrt{-g} \, \big(2\Phi\big)^{-3/2} \nonumber \\
&=& -4\varphi_0 \log{(\cos{\nu_L})}-4\varphi_0 \log{(\cos{\nu_R})}+\ldots \,.
\end{eqnarray}
The ``$\ldots$'' denotes subleading terms that do not grow in the late-time limit.
The holographic complexity thus grows linearly with time at late times 
when the electromagnetic boundary term is omitted, 
\begin{eqnarray}
\frac{d\mathcal{S}_\textrm{WDW}}{dt_{L,R}}\Big\vert_{t_{L,R}\rightarrow\infty}
&=& 8\pi \varphi_0\,T  \nonumber \\
&=&4\,S\,T+ O(T^2) \,.
\end{eqnarray}
Furthermore, the late-time 
growth rate matches the known result \eqref{eq:chargedincrement} for a near-extremal
Reissner-Nordstr\"om black hole in 3+1-dimensions, to leading order at low temperature. 


Adding a total derivative to an action does not change the equations of motion 
but can still affect the physics. Boundary terms can, for instance, 
implement a change of thermodynamic ensemble when evaluating the free energy via
the Euclidean on-shell action.  
We have demonstrated in this paper that the complexity is another physical quantity 
affected by boundary terms. 
The action of the JT theory, as it is usually written, descends from a higher-dimensional 
Einstein-Maxwell theory that inadvertently includes a boundary term that implies impermeable rather than 
permeable boundary conditions. For many purposes this is not a problem, but for calculating holographic complexity it is: we find that the CA conjecture is not consistent with the standard JT action. However,  if we remove the spurious boundary term then we get the correct answer for charged black holes and, it seems reasonable to believe, SYK.  

Fixing the electric field is tantamount to introducing an obstruction to the flow of charge, 
whereas fixing the vector potential (equivalent to fixing the chemical potential) makes the boundary permeable.
It is unphysical to impose boundary conditions on a Wheeler-DeWitt 
patch that inhibit the flow of charge across what is essentially an arbitrary internal boundary. As we have seen with the JT model, imposing an unphysical internal boundary condition gives an unphysical result for the complexity growth.

\section{Conclusion}
New physics often fixes ambiguities in the action. What was once a redundancy of the description 
becomes physically meaningful. For example, in non-gravitational physics adding a constant to the 
Lagrangian density makes no difference as the absolute zero of energy is unobservable. 
But all energy gravitates, and so in gravitational physics additive constants matter. 
Similarly, in classical physics the action is ambiguous with respect to multiplication by a constant. 
In quantum mechanics, on the other hand, multiplying the action by a constant changes $\hbar$ 
and so changes the importance of quantum effects---multiplicative constants matter. 
As a third example, quantum gravity fixes topological terms in the gravitational action, 
because the path integral includes integrals over different topologies. 

In this paper we have seen that holographic complexity fixes yet another ambiguity in the action---in holographic complexity even boundary terms that lie at unobserved internal boundaries become 
physically meaningful. Indeed, we saw that it was just such an internal boundary term that explained 
the discrepancy between the WDW action of JT gravity and the corresponding predictions of highly 
charged RN black holes. In integrating out the gauge field from RN black holes to arrive at the JT theory, 
an internal boundary term was inadvertently introduced, which is why JT gave the wrong answer for 
holographic complexity growth. For most purposes (e.g.~calculating boundary correlation functions), 
this extra term in the action does not matter---for holographic complexity, it does. 

This does raise the problem of how, given an action, we are to know whether it contains all the degrees 
of freedom necessary to describe holographic complexity. At this stage, we don't really know except to 
say that the Einstein-Maxwell Lagrangian produces plausible answers. Since these are the only massless 
degrees of freedom in the theory there doesn't seem to be much room to add anything else. 

A related open question is what principles determine the calculation of the WDW action. 
Does the ambiguity in the choice of boundary terms have to be fixed, or is it 
related to some ambiguity in the complexity? Said differently, does the WDW action with the 
``wrong'' choice of boundary terms have any complexity interpretation?

In addition to the SYK theory, there are other known holographic descriptions of charged RN black holes. 
A particularly well-known example is the D1-D5 system, and its description in terms of long multiply wound strings. 
It would be interesting to investigate the connection between such string theoretic constructions and 
the SYK model. In particular, are there approximations to the D1-D5 system whose holographic dual 
could be described by SYK? The multiply-wound nature of the D1-D5 long string allows the string to 
self-intersect and leads to an effective all-to-all coupling amongst the different windings of the string. 
This is known to be the origin of the fast scrambling nature of such systems, and is reminiscent of the 
SYK model.

To summarize, we calculated the late-time rate of change of action of the WDW patch of JT gravity. 
We found it to be zero. The CA conjecture would then predict that the rate of complexification of the 
holographic dual should be zero, but this makes no sense. Instead, we traced the discrepancy to the 
boundaries of the WDW patch, and it was there that we found the missing gates. \\

\section*{Acknowledgements}

We are grateful to Kurt Hinterbichler, Fedor Popov, Douglas Stanford, and Alexandre Streicher. We would especially like to thank Hugo Marrochio for communicating to us the results of \cite{HugoEtAl}. 
This research was supported by the John Templeton Foundation (AB), by the NDSEG program (HL), by 
NSF grant PHY-1720397 (HG), by NSF Award Number 1316699 (LS), 
by the Icelandic Research Fund under grant 163422-053 and the University of Iceland Research Fund (LT), and by the Simons Foundation (YZ). 

\begin{appendix}

\section{Not all actions are equal: The $r-\theta $ model}
\label{app:central_pot}

The sign change that we saw in the effective action for the dilaton in Section~\ref{sec:JTmodel} 
is familiar.  As a particularly simple example, consider a particle in a central potential,
\be 
L =   \frac{\dot{r}^2}{2} + r^2 \frac{\dot{\theta}^2}{2} -V(r) \,.
\label{L}
\ee
The angular momentum $l=r^2\dot{\theta}$ is conserved and the Lagrangian $L$ can be expressed as
\be 
L =   \frac{\dot{r}^2}{2} +  \frac{{l^2}}{2r^2} -V(r) \,.
\label{LL}
\ee
This equation is a numerical equality but it does not mean that we can use \eqref{LL} to get the 
equation of motion by thinking of $\frac{{l^2}}{2r^2}$ as a potential energy. It would incorrectly give the 
centrifugal force as attractive. To get the right equation you need to flip the sign of the 
$\frac{{l^2}}{2r^2}$ term,
 \be 
L' =   \frac{\dot{r}^2}{2} -  \frac{{l^2}}{2r^2} -V(r) \,.
\label{L'}
\ee
With the flipped sign, the Euler-Lagrange equation for $r$ yields the correct orbital motion 
but the actual numerical value of the action from the original Lagrangian is 
correctly given by the `wrong-sign' Lagrangian in \eqref{LL} and not by \eqref{L'}.

This sort of thing always happens when the kinetic energy of a degree of freedom 
that you integrate out becomes an effective potential energy \cite{Dyer:2008hb}. 
When deriving the JT model it happens because electric field energy is kinetic energy, 
$\vec{E}^2 = (\partial_t \vec{A})^2$, which becomes an
effective dilaton potential when the electric field is integrated out.

In Section~\ref{sec:JTmodel} we saw that the required sign-flip in the dilaton potential
can be achieved by adding a suitable boundary term to the action.
The same is true in the $r-\theta $ model.
The difference between the actions coming from \eqref{LL} and \eqref{L'}
is 
\be
I-I'=\int dt \frac{l^2}{r^2} \,,
\ee
which can be written 
\be 
I - I' = \int dt l \dot{\theta} \,.
\ee
Using the conservation of $l$ this gives
\be 
I - I' =l \int dt \dot{\theta} = l\theta|_i^f  \,,
\ee
which can be expressed as a boundary term, 
\be 
I_B= r^2\theta \dot{\theta}|_i^f \,.
\ee

Our calculations involving the JT model show that the complexity-as-action is sensitive to whether
the action is expressed in terms of kinetic energy or potential energy. In some sense, kinetic 
energy computes whereas potential energy does not. When you integrate out degrees of freedom
you may lose track of potential vs.\ kinetic energy, and so lose track of the true rate of computation 
in the system.

\section{Charged black hole thermodynamics}
\label{thermodynamics}
The free energy of a static black hole  may be obtained by continuing to Euclidean 
signature and evaluating the Euclidean on-shell action \cite{Gibbons:1976ue}. 
Which free energy this gives is determined by the boundary terms in the action. 
Let us apply this method to the Reissner-Nordstr\"om solution \eqref{RNsolution}. 

In the absence of the electromagnetic boundary term \eqref{maxwellboundaryterm} one finds 
\begin{equation}
\mathcal{S}_E= \beta F\big\vert_\mu= -S + \beta M- \beta\mu Q\,,
\label{fixed_mu}
\end{equation}
where $S=\pi r_+^2/\ell^2$ is the Bekenstein-Hawking entropy and $\mu=Q/r_+$ is the black 
hole chemical potential. This is the free energy for an ensemble where the chemical 
potential $\mu$ is kept fixed. On the other hand, when the electromagnetic boundary term is included, 
a cancellation occurs and the free energy reduces to that of a fixed charge ensemble,
\begin{equation}
\beta F\big\vert_Q= -S + \beta M\,.
\label{fixed_mu}
\end{equation}
The free energy of a two-dimensional JT black hole \eqref{2d_free_energy} 
is related to that of a near-extremal Reissner-Nordstr\"om black hole at fixed charge 
rather than at fixed chemical potential. To facilitate the comparison it is useful to 
express the Reissner-Nordstr\"om black hole mass $M$ and entropy $S$ in terms of temperature $T$ 
and (fixed) charge $Q$, 
\begin{eqnarray}
M&=&M_0+2\pi^2Q^3\ell\, T^2+\ldots\,, \nonumber \\
S&=&S_0+4\pi^2 Q^3\ell\, T+\ldots\,,
\label{RN_thermo}
\end{eqnarray}
where $M_0=Q/\ell$ and $S_0=\pi Q^2$ are the mass and entropy of an extremal black hole 
of charge $Q$ and the ``$\ldots$'' refer to higher order terms in $T$.

\section{Maxwell boundary terms in  3+1 dimensions}
 \label{sec:Maxwellboundaryterm} 

In this paper, we have seen the need to be careful about creating 1+1-dimensional Maxwell boundary 
terms when integrating down from 3+1 dimensions. However, we were able to get away without considering 
Maxwell boundary terms in the original 3+1-dimensional action, Eq.~\ref{4d_action}. This is because for 
electric RN solutions, the $F^2$ bulk action on its own requires no further boundary terms to implement 
our desired ensemble. However, even in 3+1 dimensions, there \emph{is} a time when we need to carefully 
consider Maxwell boundary terms, and that is when the black holes are not electric but magnetic. 

There is, on the face of it, a bit of a puzzle about how magnetic black holes fit into the complexity=action 
conjecture. On the one hand, one might have thought that because of electric-magnetic duality, magnetically 
charged black holes should complexify at the same rate as their electrically charged duals. On the other hand, 
since the bulk Maxwell Lagrangian changes sign under  $\vec{E} \leftrightarrow \vec{B}$, 
\begin{equation}
F_{\mu \nu} F^{\mu \nu} \sim  \vec{B}^2 -\vec{E}^2,
\end{equation}
it looks like electric and magnetic black holes cannot have the same rate of action growth. 

The resolution is  in the boundary condition for electromagnetism \cite{Brown:2015lvg}. The form of the Einstein-Maxwell action 
given in Eq.~\ref{4d_action}, with an $F^2$ bulk term and no further boundary term, is appropriate to an 
ensemble that keeps ${A}_{\mu}$ fixed on the boundary.  To keep the dual vector potential $\tilde{A}_{\mu}$ fixed at the boundary, 
as would be the ensemble appropriate for a magnetically charged black hole, we should include a boundary term. 
Adding the appropriate boundary term from Braden \emph{et al.} \cite{Braden:1990hw} 
gives the rate of action growth for a RN black hole of magnetic charge $P$ as
 \begin{equation}
\frac{dS}{d (t_L + r_R) }   = \frac{P^2 }{r_-} - \frac{P^2 }{r_+}    . 
 \end{equation}
Comparing to Eq.~\ref{eq:chargedincrement}, we see that the action growth of magnetic and 
electric black holes are related by a $Q \leftrightarrow P$ symmetry. Even though the value of 
the bulk action apparently breaks electric-magnetic duality, the duality is restored by boundary terms. 
This is analogous to the situation encountered in \cite{Hawking:1995ap,Brown:1997dm}.

\end{appendix}

\end{document}